# An Interpretable AI framework Quantifying Traditional Chinese Medicine Principles Towards Enhancing and Integrating with Modern Biomedicine

Haoran Li[1]†, Xingye Cheng[1]†, Ziyang Huang[1], Jingyuan Luo[2, 3, 4], Qianqian Xu[2, 3], Qiguang Zhao[2], Tianchen Guo[2], Yumeng Zhang[2], Linda Li-Dan Zhong[5], Zhaoxiang Bian[3, 4, 6], Leihan Tang[7]*, Aiping Lyu[2, 3, 8]*, Liang Tian[1, 8, 9]*

[1]*Department of Physics, Hong Kong Baptist University, Hong Kong, China*

[2]*School of Chinese Medicine, Hong Kong Baptist University, Hong Kong, China*

[3]*Vincent V.C. Woo Chinese Medicine Clinical Research Institute, Hong Kong Baptist University, Hong Kong, China*

[4]*Centre for Chinese Herbal Medicine Drug Development, Hong Kong Baptist University, Hong Kong, China*

[5]*School of Biological Sciences, Nanyang Technological University, Singapore*

[6]*The Chinese Medicine Hospital of Hong Kong, Hong Kong, China*

[7]*Center for Interdisciplinary Studies and Department of Physics, Westlake University, Hangzhou, China*

[8]*Institute of Systems Medicine and Health Sciences, Hong Kong Baptist University, Hong Kong, China*

[9]*Institute of Computational and Theoretical Studies, Hong Kong Baptist University, Hong Kong, China*

† These authors contributed equally.

* Correspondence and requests for materials should be addressed to Leihan Tang, Aiping Lyu or Liang Tian. (Email: tangleihan@westlake.edu.cn; aipinglu@hkbu.edu.hk; liangtian@hkbu.edu.hk)

## Abstract

Traditional Chinese Medicine diagnosis and treatment principles (TCM principles), established through centuries of trial-and-error clinical practice, directly maps patient-specific symptom patterns to personalised herbal therapies. These empirical holistic mapping principles offer valuable strategies to address remaining challenges of reductionism methodologies in modern biomedicine, such as the complexity of multifactorial diseases and the uncertainties arising from inter-individual differences and the bench-to-clinic gap. However, the lack of a quantitative framework and molecular-level evidence has limited their interpretability and reliability. Here, we present a Transformer-based autoencoder AI framework trained on ancient and classical TCM formula records to quantify the symptom pattern-herbal therapy mappings. Interestingly, we find that empirical TCM diagnosis and treatment are consistent with the encoding-decoding processes in the AI model, yielding aligned low-dimensional representations that link complex symptom patterns to intricate herb combinations. This enables us to construct an interpretable TCM embedding space (TCM-ES) using the model's quantitative representation of TCM principles. Validated through broad and extensive patient data from TCM clinics and hospitals, the TCM-ES offers universal quantification of the TCM practice and therapeutic efficacy. We further map biomedical entities (e.g., diseases, herbal compounds, drugs, and target proteins) into the TCM-ES through correspondence alignment. We find that the principal directions of the TCM-ES are significantly associated with key biological functions (such as metabolism, immune, and homeostasis), and that the disease and herb embedding proximity aligns with their genetic relationships in the human protein interactome, which demonstrate the biological significance of TCM principles. Moreover, the TCM-ES uncovers latent disease relationships, and provides alternative metric to assess clinical efficacy for modern disease-drug pairs. Finally, guided by the TCM-ES, we construct a comprehensive and integrative TCM knowledge graph, which predicts potential associations between diseases and targets, drugs, herbal compounds, and herbal therapies, providing TCM-informed opportunities for disease analysis and drug development. Together, this framework establishes quantitative foundations for advancing TCM standardization and modernization towards a valuable enhancement to modern biomedical science.

## Introduction

The development of modern biomedicine has accelerated the process of drug discovery through its systematic identification of disease mechanisms and molecular targets. In typical modern biomedical drug development, the reductionism or bottom-up paradigm [1] begins with the clinical manifestation and symptoms of a disease, followed by the identification of the underlying genetic or molecular causes of the disease, then determining the relevant targets and developing the targeted drugs aimed to restore normal biological function (Fig. 1a, upper path) [2]. While highly effective in many contexts, this paradigm faces challenges due to the increasing complexity and dimensionality along the path, which involves a wide range of interacting factors, such as genetic, molecular, environmental, etc [3,4]. Furthermore, this paradigm also suffers from the inherent uncertainties in translating early-stage *in vitro* and *in vivo* experimental findings to human applications [5,6], and in addressing the inter-individual differences in therapeutic responses [7,8].

Traditional Chinese Medicine (TCM) diagnosis and treatment principles (or TCM principles), derived through centuries of trial-and-error clinical practice, provide an alternative holistic or top-down paradigm [9–11]. Compared to the modern biomedical paradigm, TCM principles present empirical direct mappings between complex symptom patterns and various herbal therapies (Fig. 1a, lower path). In the diagnosis process, through detailed symptom descriptions of a patient's overall condition (via inspection, auscultation, inquiry, and palpation) [12], TCM applies the empirical principles to differentiate the observed symptom pattern/set into TCM syndromes, which is known as Syndrome Differentiation [13]. For example, the symptom pattern including *vomiting* and *food retention* (Fig. 1b, left) is differentiated as Cold Syndrome (Fig. 1b, middle). In the treatment process, the differentiated syndromes then guide the prescription of personalised herbal therapies based on TCM-defined herb natures and functions. For instance, Cold Syndrome is typically treated with hot/warm-natured herbs (Fig. 1b, right). A brief introduction to key TCM concepts is provided in Supplementary Information (SI) Text 1. In these two processes, syndromes essentially serve as an intermediary that links complex symptom patterns to herbal therapies [13,14]. These empirical mapping principles hold promise in reducing the mentioned complexity and uncertainty in modern biomedical paradigm through: (1) offering a holistic view of human physiology [15]; (2) enabling the design of personalised herbal therapies tailored to patient-specific symptom patterns [16]; and (3) drawing on centuries of clinical experience directly grounded in human patients [17,18].

Despite the potential of TCM principles to enhance modern biomedicine, critical research gaps hinder their integration [19]. First, the empirical mappings between symptom patterns and herbal therapies (through Syndrome Differentiation) lack quantitative foundation [20,21], making it challenging for interpretation and standardization. Additionally, the variability of herbal therapy efficacy resulting from intricate herb interactions and personalised clinical context remains poorly quantified and evaluated [22,23]. Moreover, TCM principles lack support from modern biomedical mechanisms, limiting its scientific acceptance and integration with modern biomedicine [24]. Previous network-based frameworks partially bridged these gaps by modelling disease/symptom-gene-drug/herb relationships [25–27]. However, these methods often rely on multi-layered networks and are limited to additive or pairwise relationships, which overlook TCM's nonlinear and combinatorial factors like group enhancement among symptoms [28,29] and synergistic effect of herbs [30,31]. While deep-learning models were developed to address the high-dimensionality and nonlinearity in TCM data [32,33], they often operate as "black boxes", lacking transparency and interpretability, which raises concerns about their clinical reliability. Collectively, these limitations underscore the need for an interpretable framework that quantifies the empirical knowledge of TCM for further integration with molecular-level entities towards enhancing the capabilities of modern biomedicine.

In this work, we present a Transformer-based autoencoder [34,35] AI framework to systematically quantify the TCM principles. By training the model on ancient and classical TCM formula records, we represent TCM entities as measurable vectors via an encoding-decoding process (Fig. 1c). We show that the TCM diagnosis and treatment closely align with the model's encoding-decoding process, providing a solid quantitative foundation for the TCM principles. Building on this insight, we established the TCM embedding space (TCM-ES) using the model's low-dimensional representation (**Fig. 1c**, middle), which universally quantifies the TCM practice and evaluates the clinical efficacy across multiple clinical datasets from hospitals and clinics. To further explore the biological significance of TCM principles, we incorporated biomedical entities into the TCM-ES through correspondence alignment (**Fig. 1d**). Finally, leveraging the integrated TCM-ES, we propose TCM-inspired insights to facilitate the discovery of disease-disease relationship, quantify drug efficacy, and predict novel therapeutic candidates for complex diseases, therefore demonstrating how TCM knowledge can inform and enhance modern biomedical paradigms.

## Results

### The AI model quantifying the symptom pattern-herbal therapy mappings

We developed an AI model to establish quantitative mappings between symptom patterns and herbal therapies in TCM. To achieve this, we used 84,491 ancient and classical TCM formula records (see **Methods 1.1**) for model training. These formula records, developed and continuously refined through long-term clinical practice, have been preserved through centuries of clinical application, demonstrating empirically validated therapeutic efficacy. Each of the formula records contains: (1) indicated symptom pattern; (2) TCM syndromes differentiated based on the symptom pattern; and (3) herbal compositions of the formula tailored to the syndromes. For instance, the classical formula Si-Jun-Zi-Tang [36] (**SI Table S1**) addresses the symptom pattern of *loose stools*, *decreased appetite*, *emaciation, fatigue, and shortness of breath*, which is differentiated as Spleen Qi Deficiency and Lung Qi Deficiency syndromes. Based on these syndromes, the formula comprises four herbs: Ren-Shen (*Ginseng Radix et Rhizoma*), Bai-Zhu (*Atractylodis Macrocephalae Rhizoma*), Fu-Ling (*Poria*), and Gan-Cao (*Glycyrrhizae Radix et Rhizoma*).

To ensure that the model learns objective mappings, we only used two components (the symptom pattern and herbal compositions) from each formula record for training, while masking the syndrome information. Treating these two components as therapeutic matches, we projected them into a quantitative embedding space in which matched pairs show greater proximity than unmatched pairs, thereby capturing the symptom pattern-herbal therapy mappings (**Fig. 1c**). Additionally, the embeddings were also expected to capture the symptom-symptom and herb-herb inter-relationships. To achieve these requirements, the model architecture was designed with three key schemes (**Extended Fig. E1**, Step 1): (1) An autoencoder with bottleneck dimensionality reduction to represent complex symptom patterns and herb combinations as measurable vectors; (2) A contrastive learning mechanism [37] to embed symptom patterns and formulas into the same space and enforce the spatial proximities between matched pairs; and (3) Transformer [34] blocks (in encoders and decoders) with multi-head attention mechanisms to quantify self-attentions (symptom dependencies and herb combination principles) and cross-attentions (bidirectional symptom-herb correspondence) as visualised in **Extended Fig. E2a-h**. The model was trained exclusively on the above-mentioned ancient and classical TCM formula records through self-supervised learning (see **Methods 4.1-4.3**). During training, we validate the model's plausibility by monitoring the loss on the validation set (**SI Fig. S2**) and evaluating its decoding performance on the test set (**Extended Fig. E3**). The optimal dimensionality of the bottleneck/embedding layer was examined, which captures the complexity of TCM principles while minimizing redundancy (see **Methods 4.4** and **Extended Fig. E4a**). We performed principal component analysis (PCA) on the embeddings of TCM formula records (see **Methods 5.1**), and revealed that the first three principal components (PCs) capture over 72.8% of the variance and the first six PCs account for more than 98.7% (**Extended Fig. E4b**). In contrast, PCA applied to the raw data shows highly scattered variance, with the first 500 PCs capturing less than 88.6% of the variance (**Extended Fig. E4c**). This comparison underscores the necessity of nonlinear dimensionality reduction, as it effectively captures the principal directions of the data and distils key features that cannot be achieved by conventional linear dimensionality reduction.

Using the trained model's encoders, we embedded various TCM entities, including symptoms, symptom patterns (e.g., patients' conditions or diseases), herbal therapies (e.g., herb, herb-pairs, and formulas), into the unified latent space (see **Methods 4.3** and Extended **Fig. E1**, Step 2). Within this space (**Fig. 2a**), the spatial relationship quantitatively reflects the symptom pattern-herbal therapy associations learned by the model from TCM clinical experience.

## Alignment between TCM principles and the AI encoding-decoding processes

We compared the embeddings generated by the encoding-decoding process in the AI model against TCM syndromes that were deliberately excluded during model training. Interestingly, we found that the embedding layer aligns closely with the TCM syndromes. This demonstrates that the model's data-driven encoder-decoder recapitulates TCM's diagnosis (encoding symptom patterns to syndromes) and treatment (decoding herbal therapies from syndromes) principles optimised by centuries of clinical trial-and-error. Specifically, using 7,098 formula records with TCM-differentiated syndromes, we calculated the tuning curves of the model's 256 embedding-layer neurons for 40 most frequent syndromes (See **Methods 4.5** and **Fig. 2b**, left). We organised the neurons (columns) by hierarchical clustering [38] on the tuning curves, and the syndromes (rows) by similarity of their associations with symptoms and herbs in TCM formula records (see **Methods 6.1** and **Fig. 2b**, right). The tuning curves suggest that individual neurons exhibit selective preferences for specific syndromes. Moreover, the block-diagonal pattern reveals a close correspondence between syndrome clusters and neuron clusters.

We further confirmed the alignment between the embeddings of symptom patterns and their corresponding TCM-differentiated syndromes. By visualizing the embeddings of the symptom patterns differentiated as typical TCM syndromes in the formula records (**Fig. 2c**), we observed that the model provides distinguishable and continuous representations for different symptom patterns, such as those differentiated as Cold, Heat, or Deficiency syndromes clustering along the first three PCs, respectively. Symptoms with largest projections on each PC are listed in **SI Table S3**.

We also validated the alignment between herb embeddings and their TCM-defined natures by examining their distribution along each PC. Herbs with larger projections (higher rank) on PC-1 (see **Methods 5.1**), such as Wu-Zhu-Yu (*Euodiae Fructus*) and Gao-Liang-Jiang (*Alpiniae Officinarum Rhizoma*), are mainly classified as hot/warm (**Fig. 2d**), corresponding to the Cold syndrome [39] (**Fig. 2c**). Herbs with larger projections on PC-2, such as Huang-Qin (*Scutellariae Radix*) and Zhe-Bei-Mu (*Fritillariae Thunbergii Bulbus*), are mostly categorised as cold/cool (**Fig. 2e**), aligning with the Heat syndrome (**Fig. 2c**). Herbs with higher projections on PC-3, such as Du-Zhong (*Eucommiae Cortex*) and Gou-Qi-Zi (*Lycii Fructus*), are mostly defined as neutral or warm/hot natures (**Fig. 2f**), corresponding to the Deficiency syndrome (**Fig. 2c**). Herbs with largest projections on each PC are listed in **SI Table S3**.

Collectively, the model-derived embedding, termed the TCM embedding space (TCM-ES) provides a quantitative foundation for TCM diagnosis and treatment principles. The continuous nature of the TCM-ES enables a more objective and precise representation of symptom patterns and herbal therapies beyond discrete syndrome categories. Based on this insight, we utilise the TCM-ES for the subsequent quantitative analysis.

## Universally quantifying TCM practice

The TCM-ES universally quantifies the TCM practice, including patient symptom co-occurrence, co-prescriptions of herbs, and officially indicated symptom-herb therapeutic correspondences. These relationships were validated using two extensive TCM clinical datasets: (1) 18,892 general TCM clinical cases collected from 140 TCM hospitals and clinics in mainland China over a 2-year period (see **Methods 2.1**); and (2) 150 long COVID clinical cases derived from a multicentre observational study [40] involving post-discharge patients in Hong Kong, who received three to six months of TCM treatments and follow-up visits (see **Methods 2.2**). The clinical cases represent a broad spectrum of population and conditions (**Extended Fig. E5**). Additionally, we collected 3,366 officially indicated symptom-herb pairs from the *Chinese Pharmacopoeia* [41] (see **Methods 2.4**).

We first demonstrated that symptom co-occurrence in clinical cases is quantified by the symptom-symptom embedding proximity (**Fig. 3a**, left). For each pair of symptoms, we calculated: (1) the Euclidean distances between their embeddings in the TCM-ES; and (2) their co-occurrence frequencies in clinical records measured by the Jaccard Index. For the general TCM clinical cases, a significant negative Pearson Correlation Coefficient (PCC) is observed between embedding distance and co-occurrence frequencies ($p = 4.3 \times 10^{-59}$, **Fig. 3a**, left-middle), indicating that symptom pairs embedded more proximal within the TCM-ES are more likely to co-occur. The analysis of the long COVID cases reveals a similar negative correlation ($p = 2.4 \times 10^{-5}$, **Fig. 3a**, right-middle) despite symptom homogeneity among patients within a specific disease. Furthermore, we also validated the co-occurrence pattern of disease associated symptoms across 479 human diseases (see **Methods 3.1**). Symptom pairs sharing more associated diseases exhibited greater embedding proximity ($p = 4.5 \times 10^{-33}$, **Fig. 3a**, right). We reproduced the above results using alternative metrics to measure the co-occurrence frequency (**Extended Fig. E6a**).

Similarly, we showed that herb co-prescription is quantified by the herb-herb embedding proximity (**Fig. 3b**, left). We analysed the herbal compositions of the prescribed formulas and observed significant negative correlations between pairwise herb embedding distance and co-prescription frequency (Jaccard Index) for both general TCM clinical cases ($p = 8.7 \times 10^{-264}$, **Fig. 3b**, middle) and long Covid cases ($p = 3.4 \times 10^{-34}$, **Fig. 3b**, right). It indicates that herb pairs that are co-prescribed more frequently in practice are embedded more proximal within the TCM-ES. The results were reproduced using alternative co-prescription metrics (**Extended Fig. E6b**).

We further demonstrated that the herb-symptom embedding proximity aligns with officially indicated therapeutic correspondence (**Fig. 3c**, left), using the herb-symptom pairs documented in the *Chinese Pharmacopoeia*. The result shows that indicated herb-symptom pairs exhibit significantly more proximal embeddings in the TCM-ES compared to random pairs ($p < 2.78 \times 10^{-64}$, **Fig. 3c**, right).

Evaluating the clinical efficacy of TCM herbal therapies

We next show that the TCM-ES quantifies TCM clinical therapeutic efficacy, validated using patient data from TCM hospitals and clinics.

First, we found that a formula will exhibit better efficacy to a patient's condition if their embeddings are more proximal within the TCM-ES (**Fig. 3d**, left). We analysed a well-structured clinical dataset for the COVID-19 patients with various personalised conditions [42], who received online consultation services and herbal treatments (based on nine standard formulas to ensure telemedicine safety) during the Omicron outbreak in Hong Kong (see **Methods 2.3**). For our analysis, we focused on 1,128 patients who met the following criteria: (1) complete and unified symptom information for both the initial and follow-up visits was documented; (2) prescribed one of the nine standard formulas without modifications; and (3) the time interval between the initial and follow-up visits was 5-9 days, consistent with the formulas' 5-day regimen (see **Methods 2.3**). All patients' initial conditions and the nine standard formulas were mapped into the TCM-ES based on their symptom patterns in the initial visits and the herbal compositions, respectively (**Fig. 3d**, right). Embedding Euclidean distances between each patient's initial condition and the prescribed formula were calculated. Since each symptom's severity scores were recorded for both visits, the patient's condition improvement was quantified as the total score reduction between initial and follow-up visits (see **Methods 2.3**). The analysis revealed a significant negative correlation between condition-formula embedding distances and condition improvements ($p = 3.5 \times 10^{-5}$, **Fig. 3e**), indicating that the formulas exhibit superior therapeutic effects for better-matched patients' conditions in the TCM-ES. Note that, when comparing condition improvements across patients, our results exclude placebo effects: under a null (placebo-only) model, the therapeutic differences will not correlate with embedding proximity. This analysis was replicated in the long COVID clinical dataset by comparing condition improvements between consecutive visits and the condition-formula embedding distance (**SI Fig. S7**).

We further demonstrated that the alleviated symptoms are embedded in closer proximity to the prescribed formulas (**Fig. 3f**, left). For each patient in the COVID-19 dataset, we compared the initial and follow-up symptom scores, and identified alleviated symptoms (severity scores decreased during follow-up visit) and unalleviated symptoms (severity scores unchanged or increased). The analysis suggested that improved symptoms embedded significantly closer to the prescribed formulas in the TCM-ES than those unalleviated ones (paired Student's $t$-test, $p = 1.24 \times 10^{-53}$, **Fig. 3f**, middle-left). We validated these findings in the other two clinical datasets across various diseases (the general TCM cases) and longer time spans (the long COVID cases with multiple visits) by identifying alleviated symptoms (see **Methods 2.1-2.2**). The results consistently reveal significantly greater proximity from patients' prescribed formulas to their alleviated symptoms in the general TCM clinical cases ($p = 1.39 \times 10^{-247}$, **Fig. 3f**, middle-right) and the long COVID cases ($p = 3.87 \times 10^{-14}$, **Fig. 3f**, right). Note that, due to the subtle progression of long COVID, follow-up records typically focused on significantly improved symptoms while neglecting partial changes, we thus compared embedding distances between herbal formulas and improved versus all other symptoms across patients. Notably, the formula-symptom pairs in the two datasets above exhibit shorter embedding distances than the COVID-19 dataset, as the formulas in these datasets were more diverse (**SI Fig. S6**) and customised/personalised to patients' conditions, unlike the standard formulas (for telemedicine safety) in the COVID-19 dataset.

Finally, we showed that the TCM-ES partially quantifies the synergetic (enhanced) or antagonistic (reduced) therapeutic efficacy of co-prescribed herbs. The difference in embedding distance from a targeted symptom ($S$) to individual herbs ($H_A, H_B$) versus their co-prescription (herb-pair $H_{AB}$) in the TCM-ES indicates the herbal combinatorial effects (**Fig. 3g**, left). We focused on 349 prevalent and relevant herb-herb-symptom triplets ($[H_A, H_B, S]$) and statistically evaluated their clinical synergistic and antagonistic effects in the general TCM clinical dataset (see **Methods 6.4**). We observed a negative PCC ($p = 0.026$, **Fig. 3g**, right) between the difference in embedding distance to the symptom (herb-pair versus individual herb) and the efficacy difference. This suggests that herb pairs embedded closer to symptoms than individual herbs exhibit a synergistic effect, whereas herb pairs farther away show a potential antagonistic effect.

## Mapping biomedical entities into the TCM-ES

By keeping the embedding of TCM entities unchanged, we mapped modern biomedical entities into the TCM-ES using correspondence alignment (**Figs. 1d** and **4a**; **Extended Fig. E1**), including: (1) Human diseases [26,43] embedded through associated symptom patterns [26] (**Extended Fig. E1, step 3**; **Methods 3.1** and **4.6**); (2) Herbal compounds and target proteins (from the HIT 2.0 database [44]) embedded based on their corresponding herbs' embeddings (**Extended Fig. E1, step 3**; **Methods 3.2** and **4.6**); and (3) FDA-approved modern drugs [45] embedded based on their chemical structure similarity [46] to already embedded herbal compounds (**Extended Fig. E1, step 3**; **Methods 3.5** and **4.7**). Note that the drugs were filtered according to their compatibility with the encoder trained on TCM herbal compounds to ensure embedding accuracy (see **Methods 4.7**).

Note that conventional TCM paradigm lacks molecular-level insights, leading to the exclusion of biomedical entities (diseases, compounds, targets, genes, etc.) from model training and TCM-ES construction. Therefore, these entities' embeddings rely solely on their associations with symptoms or herbs in the TCM-ES.

## Biological significance of the TCM-ES

The embedding of biomedical entities (**Fig. 4b**) enables investigation of biological functions associated with the TCM-ES and validating molecular plausibility of herbal therapies.

We first demonstrated that the principal directions (i.e., the PCs) of the TCM-ES are significantly associated with key biological functions. For each PC of the TCM-ES, we identified the top $K_t = 200$ target proteins with the highest projections and then performed Enrichment Analysis (EA) [47,48], revealing distinct biological processes (see **Methods 5.2**). The PC-1 showed dominant associations with metabolic processes, including the reactive oxygen species metabolic process (GO:0072593, $p < 2.2 \times 10^{-16}$) and fatty acid metabolic process (GO:0006631, $p = 2.20 \times 10^{-16}$), where metabolism-related proteins (i.e., proteins involved in metabolic processes, see **Methods 5.2**) exhibit significantly higher projection ranks (see **Methods 5.1**) than non-metabolic ones (**Fig. 4b**, PC-1 inset). The PC-2 demonstrated strong associations with immune-related processes, including immune response-regulating signalling pathway (GO:0002764, $p < 2.2 \times 10^{-16}$) and T cell activation (GO:0042110, $p < 2.2 \times 10^{-16}$), with immune-related proteins showing significantly higher projection ranks (**Fig. 4b**, PC-2 inset). The PC-3 exhibited associations with homeostatic processes, such as lipid homeostasis (GO:0055088, $p < 2.2 \times 10^{-16}$) and multicellular organismal homeostasis (GO:0048871, $p = 1.21 \times 10^{-13}$), where homeostasis-associated proteins displayed significantly higher ranks (**Fig. 4b**, PC-3 inset). Detailed EA results were presented in the **SI Table S8**. The EA was conducted with varying $K_t$ and the results revealed similar biological functions.

We then assessed the biological plausibility of human disease (symptom pattern) embeddings and herb embeddings, by comparing their spatial relationships within the TCM-ES and their genetic relationships (**Figs. 4c-e**, left panels) on the human protein-protein interaction (PPI; see **Methods 3.4**) network [49]. The PPI distance among diseases and herbs was measured by the module separation between herb target proteins or disease-associated genes [50] (see **Methods 6.5**). First, we demonstrated that TCM understanding of diseases through symptom patterns aligns with their genetic relationships. By analysing 79 diseases with ≥ 20 identified associated genes [50] (see **Methods 3.3**), we observed significant positive correlations between disease-disease proximity within the TCM-ES and on the PPI network: diseases embedded closer in the TCM-ES tend to have associated gene modules closer on the PPI network ($p = 2.1 \times 10^{-30}$, **Fig. 4c**). Second, we show that frequently co-prescribed herbs exhibit more shared targets: herbs embedded closer to each other in the TCM-ES tend to target more overlapped protein modules ($p = 2.1 \times 10^{-59}$, **Fig. 4d**). Additionally, herb-disease embedding distances in the TCM-ES correlate with their genetic relationships: herbs with shorter embedding distances to specific diseases in TCM-ES demonstrated significantly smaller PPI distance between their targets and disease genes ($p = 1.5 \times 10^{-10}$, **Fig. 4e**). We replicated the analyses about diseases with refined disease associated symptoms based on Mayo Clinic [51] data (**SI Text 5**) and observed the consistent pattern (**SI Fig. S9**).

We further assessed TCM-derived embeddings' ability to reconstruct verified biomedical relationships. We found that target proteins embedded closer to each other in the TCM-ES tend to have shorter shortest-path lengths on the PPI network (**Extended Fig. E7a, b**). Note that proteins frequently co-targeted by the same herb show no necessary PPI proximity (**Extended Fig. E7c**), confirming that the protein-protein relationships were inferred from TCM herb-herb embedding proximity.

Collectively, despite the absence of molecular or genetic knowledge within TCM principles, the TCM-ES analysis revealed that TCM principles partially aligned with latent biomedical mechanisms and reproduced biomedically plausible associations verified in modern biomedical paradigms.

## Uncovering latent pathogenic relationships between diseases from TCM perspectives

Building on the abovementioned biologically meaningful symptom-based representations of diseases, we found that the TCM-ES quantitatively uncovers latent disease relationships.

To systematically analyse the disease relationships within the TCM-ES, we constructed a *K*-nearest neighbours (KNN) network for the 79 diseases (each with ≥ 20 associated genes), connecting each disease to its $K_d = 7$ nearest neighbours, retaining edges where both diseases ranked within each other's top $K_d$ neighbours (**Fig. 5a**). The network revealed meaningful clusters, where diseases sharing MeSH categories [52] (i.e., with similar pathogenic mechanisms, see **Methods 3.1**) grouped together. For example, disorders related to the gut formed a distinct cluster (**Fig. 5a**, left). Quantitative analysis confirmed significantly shorter embedding distances between diseases sharing MeSH categories compared to unrelated pairs (paired Student's *t*-test, $p = 4.15 \times 10^{-42}$, **Fig. 5b**). We confirmed the robustness of the network topologies by varying $K_d$ (**SI Fig. S10**).

The disease network also revealed latent connections between diseases without shared genes or symptoms (**Fig. 5a**, red edges), which aligned with established biomedical evidence. For example, a cluster (**Fig. 5a**, blue area) containing neurodegenerative diseases (e.g., *Amyotrophic Lateral Sclerosis* and *Parkinson's Disease)* and retinal degenerative diseases (e.g., *Macular Degeneration* and *Retinitis Pigmentosa*) exhibited shared pathogenesis: *Amyotrophic Lateral Sclerosis* and age-related *Macular Degeneration* share genetic determinants [53], and retinal changes serve as early biomarkers for neurodegenerative disorders [54], with oxidative stress and mitochondrial dysfunction playing central roles in the pathogenesis of both types of diseases [55]. The network identified another cluster (**Fig. 5a**, yellow area) linking cardiovascular diseases (e.g., *Cardiac Arrhythmias* and *Cardiomyopathy*) and neurological disorders (e.g., *Peripheral Nervous System Diseases* and *Cerebellar Ataxia*). Mechanistic evidence supports the causal association: *Arteriosclerosis*-induced cerebral hypoperfusion elevates neurological disorder risks [56], and central nervous system disorders can induce cardiovascular diseases mediated by autonomic nervous system [57].

Notably, we confirmed the proximity of these diseases in TCM-ES was not due to trivial semantic similarity of their associated symptoms. The attention maps revealed that different diseases exhibited distinct symptom focus (**Extended Fig. E2i-k**). Their embeddings were determined by unique symptom dependency patterns, rather than by a simple aggregation of symptom semantics.

# TCM-ES provides an alternative metric for assessing clinical efficacy of modern drugs

We next demonstrated the TCM-ES infers clinical efficacy of drug-disease pairs through embedding proximity.

We used a dataset [45] on drug-disease indication pairs and their efficacy. This dataset includes 402 indicated drug-disease pairs (238 drugs, 78 diseases) with solid literature evidence (see **Methods 3.5**). Among these pairs, efficacy of 204 pairs was quantified using Relative Efficacy (RE) scores [45] based on adverse event report from FDA [58], ranging from 0 (lowest) to 1 (highest). For our analysis, we further filtered these pairs to ensure compatibility with the TCM-ES, retaining only diseases with available symptom data and drugs that could be accurately decoded by the TCM-ES (**Methods 4.7**). This resulted in 125 drug-disease pairs (94 drugs, 27 diseases), of which 77 had available RE scores.

The drugs and diseases were mapped into the TCM-ES respectively (**Extended Fig. E8a**). Indicated drug-disease pairs show shorter embedding distances than random pairs (left-skewed $z$-score distribution, **Extended Fig. E8b**), suggesting that the TCM-ES captures the drug-disease therapeutic correspondence.

We next used the TCM-ES to evaluate the clinical efficacy of these drug-disease pairs. As the distribution of drugs and diseases in the TCM-ES was uneven (e.g., *Cardiac Arrhythmias* was close to most drugs, while *Rheumatoid Arthritis* was more distant overall, **Extended Fig. E8a**), relying on absolute distances alone fails to capture the two-way relative proximity between drug and disease. To address this, we introduced the bi-directional $z$-score (BZS) for each drug $x$ and disease $y$ pair:

$$BZS(x, y) = (z_{X \to Y}(x, y) + z_{Y \to X}(x, y))/2,$$

which combined the drug-to-disease $z$-score (quantifying how specifically a drug targets a disease) and the disease-to-drug $z$-score (indicating how proximal a disease is embedded to that drug). This metric provides a more balanced assessment, where smaller BZS indicates better mutual-matching within the TCM-ES (see **Methods 6.6**). It is found that BZS exhibits a significant negative correlation with RE scores ($p = 0.0012$, **Fig. 5c**), indicating that better mutual-matching in the TCM-ES corresponded to relatively higher therapeutic efficacy. Robustness of this pattern was confirmed across datasets with different drug screening criteria, refined disease embeddings based on symptoms from Mayo Clinic, and up-to-date FDA adverse report data (**SI Text 7** and **Fig. S12**).

Additionally, we illustrated that the TCM-ES presents a quantitative characterization of cross-disease drug efficacy. For diseases with at least three drugs having RE scores, a general trend was observed: diseases with smaller BZS to their corresponding drugs tend to achieve higher RE scores and vice versa. (**Fig. 5d**). Diseases like *Cardiac Arrhythmias* [59,60] exhibited close embedding proximity to indicated drugs, indicating precise therapeutic targeting and high efficacy. Conversely, complex diseases with diverse symptom patterns and subtypes, such as *Rheumatoid Arthritis* [61], displayed embedding separation with indicated drugs and lower efficacy. Note that, although this evaluation metric is limited by the information contained in the drug structural representation and the level of detail in the disease associated symptoms, we still observe a clear correlation.

## A TCM-informed knowledge graph for predicting potential therapeutic candidates

Leveraging the capacity of the TCM-ES, we constructed a comprehensive and integrative TCM knowledge graph, through which we made predictions of therapeutic candidates to facilitate TCM-informed drug development for complex diseases (**Fig. 6**). The clinical grounded knowledge graph centres on symptom-based disease embeddings and links each disease to its nearest neighbouring entities across four types of therapeutic candidates (see **Methods 6.7**), including proteins (therapeutic targets), drugs (repurposing candidates), herbal compounds (novel therapeutic agents), and herbal therapies (conventional remedies).

To illustrate the predictions, we presented the predicted therapeutic candidates for *Rheumatoid Arthritis* (*RA*). As the limited symptom terms (derived from MeSH [43]) led to an oversimplified description of *RA*, we incorporated detailed TCM symptoms to embed *RA* [62] for more precise embedding (**SI Text 8**). The TCM-ES successfully distinguishes two clinical recognised subtypes of *RA* (Cold and Heat) [62–64], with associated genes [62] of each subtype forming clusters around their respective subtypes (**Extended Fig. E9**), demonstrating the model's capacity to capture the clinical heterogeneity through symptom-based embeddings. Notably, when *RA* is embedded using literature-derived symptoms (i.e., a general and coarse symptom representation of *RA*, **Extended Fig. E8a**), some anti-inflammatory drugs indicated for *RA* (e.g., Ibuprofen) were positioned distantly from *RA*. In contrast, these drugs show closer proximity to the Heat subtype of *RA* embedded with detailed TCM symptoms (**Extended Fig. E9**). This indicates that the TCM-ES predicts these drugs exhibit enhanced anti-inflammatory effects specific to the Heat subtype, which is supported by previous research [65,66]. This observation underscores the critical need for more precise symptom characterization to advance precision medicine.

The knowledge graph predicts potential therapeutic targets and drugs for *RA*, some of which are supported by biological evidence (**Extended Table E1**). For instance, the predicted target *BTRC* (rank 3, $z = -3.20$) has been validated for its relevance to *RA*, with study showing that downregulated miR-10a in the fibroblast-like synoviocytes (FLSs) of *RA* patients accelerates IκB degradation and NF-κB activation by targeting *BTRC* [67]. Similarly, for the predicted target *MC1R* (rank 4, $z = -2.78$), prior research demonstrated its broader therapeutic potential for systemic sclerosis, neuroinflammation, *RA*, and fibrosis [68]. Among the drugs predicted for repurposing, *Arbutin* (rank 3, $z = -1.95$), originally approved for skin diseases, is suggested to be effective against *RA*. Prior study supports this prediction, showing that *Arbutin* significantly reduces CFA-induced arthritis in rats by modulating both anti-inflammatory cytokines and pro-inflammatory markers while improving oxidative biomarkers [69].

The TCM-ES further uncovers novel herbal compounds and traditional remedies with plausible therapeutic potential, some of which have been supported by existing studies **(Extended Table E1)**. For example, the predicted compound *Selenomethionine* (rank 1 in both herbal compound and drug lists, $z = -3.30$), a bioactive ingredient of Sang-Ji-Sheng (*Taxilli Herba*), has been confirmed to delay *RA* onset and reduce disease severity in mouse models [70]. At the herbal therapy level, the predicted herb Du-Zhong (*Eucommiae Cortex*, rank 1, $z = -2.81$), has been supported by previous research demonstrating the anti-*RA* activity and the pharmacokinetic profiles of its active ingredients, such as aucubin and pinoresinol glucopyranoside [71]. Moreover, the classical TCM formula *Du-Huo-Ji-Sheng-Tang* (rank 2, $z = -2.63$) has demonstrated potential in managing *RA* symptoms by significantly reducing joint inflammation and cartilage degradation in the mouse model [72].

## Discussion

By developing an AI-based framework trained on TCM formula records, we demonstrate for the first time that the TCM diagnosis and treatment processes align with the AI encoding and decoding process, providing a quantitative foundation for TCM principles. Building on this understanding, we construct the TCM-ES to effectively facilitate the quantification of TCM practice and clinical efficacy. By systematically integrating biomedical entities, we demonstrate the biological significance of TCM principles and discover TCM-informed new knowledge. Importantly, the predictive knowledge graph is grounded in the biological plausible spatial relationships within the TCM-ES, presenting TCM-informed predictions.

Our framework establishes the foundation for a series of future research directions in this promising field. First, the TCM-ES demonstrates strong potential for advancing personalised and precision medicine. In the analysis of COVID-19 cases, it effectively captured heterogeneous symptom patterns among patients with the same disease and successfully quantified therapeutic outcomes based on the embedding proximity between patients' initial conditions and the prescribed formulas. This provides a quantitative foundation for future design of personalised herbal therapies tailored to individual profiles. Furthermore, in the analysis of chronic conditions such as long COVID, where symptom alleviations were gradual, the TCM-ES still identified meaningful correspondences between herbal therapies and symptom changes. This framework enables the future work of precision herbal therapy targeting specific symptoms. Incorporating additional patient-specific information [73], such as BMI, comorbidities, age, and gender, into the TCM-ES could further enhance the personalised therapy design.

Second, our results also established a robust foundation for downstream predictive tasks. The current model already decodes (predicts) herbal therapies directly from symptom embeddings (**Extended Fig. E3**). Using similar model architecture, In the future, the unified TCM-ES can be extended to decode additional relationships (disease to herbal compounds, disease to drugs, disease to proteins, etc.) beyond simple distance-based inference. Moreover, these embeddings can serve as input features for downstream models aimed at predicting molecular interactions and drug-drug synergistic effects.

Additionally, the TCM-ES can be further refined by incorporating more detailed information about herbal therapies, such as dosage, preparation methods, and the bioactive compounds within each formula. Finally, based on these foundations, human diseases and therapy can be redefined from a dynamic TCM perspective: by integrating temporal information, diseases may be conceptualised as evolving trajectories within this multidimensional space, shaped by biomedical factors, individual variability, and clinical contexts, rather than as static entities. This more detailed characterization of patient stages enables more accurate disease classification, deeper mechanistic understanding, refined disease-progression modelling, and improved therapy design and outcome prediction.

While our framework effectively quantifies the correspondence and efficacy between TCM symptom patterns and herbal therapies, the majority of these associations remain end-to-end and do not explicitly illustrate the intermediate biological mechanisms (e.g., how herbal compounds act on targets and regulate biological functions at the pathway level). This is constrained by TCM's conventional focus on empirical end-to-end practices, which often lack molecular-level exploration such as randomised controlled trials (RCTs). However, validations on real-world clinical datasets confirmed the reliability of these end-to-end quantifications. Additionally, the integrating modern biomedical entities has demonstrated their biological plausibility, such as the functional enrichment of target proteins along PCs of the TCM-ES (Fig. 4b) and the PPI proximity between herb targets and disease-associated genes (Fig. 4e). These findings suggest that the end-to-end associations may reflect underlying complex molecular interactions, even if the mechanistic links have not yet been fully elucidated, which provides promising candidates for future experimental validation. To further investigate these intermediates, we aim to incorporate RCT-derived metrics and omics data [74] (gene expression, metabolism, microbiota, etc.) into the TCM-ES and align them with the quantitative representations of patient condition changes. Such efforts would greatly improve the mechanistic interpretability and clinical reliability of this framework, and facilitate its integration with modern biomedicine.

# Main Reference

1. Ahn, A. C., Tewari, M., Poon, C.-S. & Phillips, R. S. The limits of reductionism in medicine: could systems biology offer an alternative? *PLoS Med* **3**, e208 (2006).

2. Dugger, S. A., Platt, A. & Goldstein, D. B. Drug development in the era of precision medicine. *Nat Rev Drug Discov* **17**, 183–196 (2018).

3. Marti, A., Martinez-González, M. A. & Martinez, J. A. Interaction between genes and lifestyle factors on obesity: Nutrition Society Silver Medal Lecture. *Proceedings of the Nutrition Society* **67**, 1–8 (2008).

4. Hu, J. X., Thomas, C. E. & Brunak, S. Network biology concepts in complex disease comorbidities. *Nat Rev Genet* **17**, 615–629 (2016).

5. Johnson, J. I. et al. Relationships between drug activity in NCI preclinical in vitro and in vivo models and early clinical trials. *Br J Cancer* **84**, 1424–1431 (2001).

6. Jucker, M. The benefits and limitations of animal models for translational research in neurodegenerative diseases. *Nat Med* **16**, 1210–1214 (2010).

7. Evans, W. E. & Johnson, J. A. Pharmacogenomics: The Inherited Basis for Interindividual Differences in Drug Response. *Annu Rev Genomics Hum Genet* **2**, 9–39 (2001).

8. Jithesh, P. V. et al. A population study of clinically actionable genetic variation affecting drug response from the Middle East. *NPJ Genom Med* **7**, 10 (2022).

9. Tu, Y. The discovery of artemisinin (qinghaosu) and gifts from Chinese medicine. *Nature Medicine* vol. 17 1217–1220 Preprint at https://doi.org/10.1038/nm.2471 (2011).

10. Hao, P. et al. Traditional Chinese medicine for cardiovascular disease: evidence and potential mechanisms. *J Am Coll Cardiol* **69**, 2952–2966 (2017).

11. Hu, K. et al. Efficacy and safety of Lianhuaqingwen capsules, a repurposed Chinese herb, in patients with coronavirus disease 2019: A multicenter, prospective, randomized controlled trial. *Phytomedicine* **85**, 153242 (2021).

12. Tian, D. et al. A review of traditional Chinese medicine diagnosis using machine learning: Inspection, auscultation-olfaction, inquiry, and palpation. *Comput Biol Med* **170**, 108074 (2024).

13. Jiang, M. et al. Syndrome differentiation in modern research of traditional Chinese medicine. *J Ethnopharmacol* **140**, 634–642 (2012).

14. Lu, A., Jiang, M., Zhang, C. & Chan, K. An integrative approach of linking traditional Chinese medicine pattern classification and biomedicine diagnosis. *J Ethnopharmacol* **141**, 549–556 (2012).

15. Sun, D. et al. Differences in the origin of philosophy between Chinese medicine and western medicine: Exploration of the holistic advantages of Chinese medicine. *Chin J Integr Med* **19**, 706–711 (2013).

16. Wang, S., Long, S. & Wu, W. Application of Traditional Chinese Medicines as Personalized Therapy in Human Cancers. *Am J Chin Med (Gard City N Y)* **46**, 953–970 (2018).

17. Zhou, X. *et al.* Development of traditional Chinese medicine clinical data warehouse for medical knowledge discovery and decision support. *Artif Intell Med* **48**, 139–152 (2010).

18. Matos, L. C., Machado, J. P., Monteiro, F. J. & Greten, H. J. Understanding traditional Chinese medicine therapeutics: an overview of the basics and clinical applications. in *Healthcare* vol. 9 257 (2021).

19. Matos, L. C., Machado, J. P., Monteiro, F. J. & Greten, H. J. Can traditional chinese medicine diagnosis be parameterized and standardized? A narrative review. *Healthcare (Switzerland)* vol. 9 Preprint at https://doi.org/10.3390/healthcare9020177 (2021).

20. Tang, H., Huang, W., Ma, J. & Liu, L. SWOT analysis and revelation in traditional Chinese medicine internationalization. *Chin Med* **13**, 5 (2018).

21. Guo, Y. *et al.* Acceptability of Traditional Chinese medicine in Chinese people based on 10-year's real world study with mutiple big data mining. *Front Public Health* **9**, 811730 (2022).

22. Venkataramanan, R., Komoroski, B. & Strom, S. In vitro and in vivo assessment of herb drug interactions. *Life Sci* **78**, 2105–2115 (2006).

23. Zhou, X., Li, C.-G., Chang, D. & Bensoussan, A. Current Status and Major Challenges to the Safety and Efficacy Presented by Chinese Herbal Medicine. *Medicines* **6**, (2019).

24. Huang, N. *et al.* Possible opportunities and challenges for traditional Chinese medicine research in 2035. *Front Pharmacol* **Volume 15-2024**, (2024).

25. Gan, X. *et al. Network Medicine Framework Reveals Generic Herb-Symptom Effectiveness of Traditional Chinese Medicine.* https://www.science.org (2023).

26. Zhou, X., Menche, J., Barabási, A. L. & Sharma, A. Human symptoms-disease network. *Nat Commun* **5**, (2014).

27. Goh, K. Il *et al.* The human disease network. *Proc Natl Acad Sci U S A* **104**, 8685–8690 (2007).

28. Asmundson, G. J. G. & Katz, J. Understanding the co-occurrence of anxiety disorders and chronic pain: state-of-the-art. *Depress Anxiety* **26**, 888–901 (2009).

29. Laird, B. J. A. *et al.* Pain, Depression, and Fatigue as a Symptom Cluster in Advanced Cancer. *J Pain Symptom Manage* **42**, 1–11 (2011).

30. Zhou, X. *et al.* Synergistic Effects of Chinese Herbal Medicine: A Comprehensive Review of Methodology and Current Research. *Front Pharmacol* **Volume 7-2016**, (2016).

31. Cai, C. *et al.* Synergistic Effect of Compounds from a Chinese Herb: Compatibility and Dose Optimization of Compounds from N-Butanol Extract of Ipomoea stolonifera. *Sci Rep* **6**, 27014 (2016).

32. Zhang, H. *et al.* Transformer- And Generative Adversarial Network–Based Inpatient Traditional Chinese Medicine Prescription Recommendation: Development Study. *JMIR Med Inform* **10**, (2022).

33. Yang, G., Liu, X., Shi, J., Wang, Z. & Wang, G. TCM-GPT: Efficient pre-training of large language models for domain adaptation in Traditional Chinese Medicine. *Computer Methods and Programs in Biomedicine Update* **6**, 100158 (2024).

34. Vaswani, A. *et al.* Attention is All you Need. in *Advances in Neural Information Processing Systems* (eds. Guyon, I. et al.) vol. 30 (Curran Associates, Inc., 2017).

35. Hinton, G. E. & Salakhutdinov, R. R. Reducing the Dimensionality of Data with Neural Networks. *Science (1979)* **313**, 504–507 (2006).

36. Hong Kong Baptist University School of Chinese Medicine. Four Gentlemen Decoction. *Chinese Medicine Formulae Images Database* https://sys01.lib.hkbu.edu.hk/cmed/cmfid/detail.php?id=F00067 (2022).

37. Chen, T., Kornblith, S., Norouzi, M. & Hinton, G. A Simple Framework for Contrastive Learning of Visual Representations. in *Proceedings of the 37th International Conference on Machine Learning* (eds. III, H. D. & Singh, A.) vol. 119 1597–1607 (PMLR, 2020).

38. Jain, A. K., Murty, M. N. & Flynn, P. J. Data clustering: a review. *ACM Comput. Surv.* **31**, 264–323 (1999).

39. Liu, J., Feng, W. & Peng, C. A Song of Ice and Fire: Cold and Hot Properties of Traditional Chinese Medicines. *Front Pharmacol* **11**, 1–20 (2021).

40. Zhong, L. L. D. *et al.* Effects of Chinese medicine for COVID-19 rehabilitation: a multicenter observational study. *Chinese Medicine (United Kingdom)* **17**, (2022).

41. Chinese Pharmacopoeia Commission. *Pharmacopoeia of the People's Republic of China, 2020 Edition, Volume I.* vol. 1 (China Medical Science Press, Beijing, 2020).

42. Luo, J. *et al.* Prevalence and risk factors of long COVID 6–12 months after infection with the Omicron variant among nonhospitalized patients in Hong Kong. *J Med Virol* **95**, (2023).

43. National Library of Medicine. Medical Subject Headings (MeSH). *U.S. National Library of Medicine* https://www.ncbi.nlm.nih.gov/mesh (2022).

44. Yan, D. *et al.* HIT 2.0: An enhanced platform for Herbal Ingredients' Targets. *Nucleic Acids Res* **50**, D1238–D1243 (2022).

45. Guney, E., Menche, J., Vidal, M. & Barábasi, A. L. Network-based in silico drug efficacy screening. *Nat Commun* **7**, (2016).

46. Weininger, D. SMILES, a Chemical Language and Information System. 1. Introduction to Methodology. *J Chem Inf Model* **28**, 31–36 (1988).

47. Subramanian, A. *et al.* Gene set enrichment analysis: A knowledge-based approach for interpreting genome-wide expression profiles. *Proceedings of the National Academy of Sciences* **102**, 15545–15550 (2005).

48. Wang, J., Duncan, D., Shi, Z. & Zhang, B. WEB-based GEne SeT AnaLysis Toolkit (WebGestalt): update 2013. *Nucleic Acids Res* **41**, W77–W83 (2013).

49. Morselli Gysi, D. *et al.* Network medicine framework for identifying drug-repurposing opportunities for COVID-19. *Proceedings of the National Academy of Sciences* **118**, e2025581118 (2021).

50. Menche, J. *et al.* Uncovering disease-disease relationships through the incomplete interactome. *Science (1979)* **347**, 1257601 (2015).

51. Mayo Clinic. Diseases and Conditions. *Mayo Foundation for Medical Education and Research* https://www.mayoclinic.org/diseases-conditions.

52. National Library of Medicine. MeSH Tree Structures. *U.S. National Library of Medicine* https://www.nlm.nih.gov/mesh/intro_trees.html (2023).

53. Strafella, C. *et al.* Genetic Determinants Highlight the Existence of Shared Etiopathogenetic Mechanisms Characterizing Age-Related Macular Degeneration and Neurodegenerative Disorders. *Front Neurol* **Volume 12-2021**, (2021).

54. Normando, E. M. *et al.* The retina as an early biomarker of neurodegeneration in a rotenone-induced model of Parkinson's disease: Evidence for a neuroprotective effect of rosiglitazone in the eye and brain. *Acta Neuropathol Commun* **4**, (2016).

55. Catalani, E., Brunetti, K., Del Quondam, S. & Cervia, D. Targeting Mitochondrial Dysfunction and Oxidative Stress to Prevent the Neurodegeneration of Retinal Ganglion Cells. *Antioxidants* vol. 12 Preprint at https://doi.org/10.3390/antiox12112011 (2023).

56. Wardlaw, J. M., Smith, C. & Dichgans, M. Small vessel disease: mechanisms and clinical implications. *Lancet Neurol* **18**, 684–696 (2019).

57. Hu, J.-R., Abdullah, A., Nanna, M. G. & Soufer, R. The Brain–Heart Axis: Neuroinflammatory Interactions in Cardiovascular Disease. *Curr Cardiol Rep* **25**, 1745–1758 (2023).

58. U.S. Food and Drug Administration. openFDA Application Programming Interface. *U.S. Food and Drug Administration* https://open.fda.gov/apis/ (2023).

59. Mayo Clinic. Heart arrhythmia. *Mayo Clinic* https://www.mayoclinic.org/diseases-conditions/heart-arrhythmia/symptoms-causes/syc-20350668 (2023).

60. Schwartz, P. J. *et al.* Inherited cardiac arrhythmias. *Nat Rev Dis Primers* **6**, 58 (2020).

61. Radu, A.-F. & Bungau, S. G. Management of Rheumatoid Arthritis: An Overview. *Cells* **10**, (2021).

62. Jiang, M. *et al.* Correlation between cold and hot pattern in traditional Chinese medicine and gene expression profiles in rheumatoid arthritis. *Front Med* **5**, 219–228 (2011).

63. Lu, C. *et al.* Network-Based Gene Expression Biomarkers for Cold and Heat Patterns of Rheumatoid Arthritis in Traditional Chinese Medicine. *Evidence-Based Complementary and Alternative Medicine* **2012**, 203043 (2012).

64. Wang, W., Guan, J., Li, Z. & Wang, X. Rheumatoid arthritis characteristics and classification of heat and cold patterns–an observational study. *Heliyon* **9**, (2023).

65. Jasani, M. K., Downie, W. W., Samuels, B. M. & Buchanan, W. W. Ibuprofen in rheumatoid arthritis. Clinical study of analgesic and anti-inflammatory activity. *Ann Rheum Dis* **27**, 457 (1968).

66. Grennan, D. M., Ferry, D. G., Ashworth, M. E., Kenny, R. E. & Mackinnon, M. The aspirin-ibuprofen interaction in rheumatoid arthritis. *Br J Clin Pharmacol* **8**, 497–503 (1979).

67. Mu, N. *et al.* A novel NF-κB/YY1/microRNA-10a regulatory circuit in fibroblast-like synoviocytes regulates inflammation in rheumatoid arthritis. *Sci Rep* **6**, 20059 (2016).

68. Mun, Y., Kim, W. & Shin, D. Melanocortin 1 receptor (MC1R): Pharmacological and therapeutic aspects. *Int J Mol Sci* **24**, 12152 (2023).

69. Sial, N. T., Malik, A., Iqbal, U. & Rehman, M. F. U. Arbutin attenuates CFA-induced arthritis by modulating expression levels of 5-LOX, NF-κB, IL-17, PGE-2 and TNF-α. *Inflammopharmacology* **32**, 2377–2394 (2024).

70. Qin, J. *et al.* Supranutritional selenium suppresses ROS-induced generation of RANKL-expressing osteoclastogenic CD4+ T cells and ameliorates rheumatoid arthritis. *Clin Transl Immunology* **10**, e1338 (2021).

71. Yang, R. *et al.* Pharmacokinetics, anti-rheumatoid arthritis activity, and active ingredient contents of Eucommia ulmoides Oliv. *Fitoterapia* **170**, 105667 (2023).

72. Chen, W. *et al.* Du Huo Ji Sheng Tang inhibits Notch1 signaling and subsequent NLRP3 activation to alleviate cartilage degradation in KOA mice. *Chin Med* **18**, 80 (2023).

73. Moor, M. *et al.* Foundation models for generalist medical artificial intelligence. *Nature* **616**, 259–265 (2023).

74. Watanabe, K. *et al.* Multiomic signatures of body mass index identify heterogeneous health phenotypes and responses to a lifestyle intervention. *Nat Med* **29**, 996–1008 (2023).

## Acknowledgements

The work is supported in part by the Research Grants Council of Hong Kong (C2005-22Y) and the Hong Kong Chinese Medicine Development Fund (22B2/049A). We thank Yang-Yu Liu, Jade Shi, Yihang Wan, Qiufu Ma, Jianyang Zeng, and Xuezhong Zhou for valuable discussions or careful readings of the manuscript.

## Author Contributions

L.T., AP.L., and LH.T. conceived and designed the study; L.T. supervised the overall project; HR.L. and XY.C. performed the research; HR.L. led the model development, training, and analysis; XY.C. led the processing of TCM, biomedical, and clinical datasets; ZY.H. contributed to biomedical data processing; QQ.X., QG.Z., TC.G., and YM.Z. contributed to the manual processing of TCM data; AP.L. led the recruitment and collection of general patient clinical data across TCM hospitals and clinics in mainland China; LD.Z. collected the long COVID patient clinical data from TCM clinics in Hong Kong; JY.L. and ZX.B. led the recruitment and collection of the COVID-19 patient clinical data in HKBU TCM telemedicine centre; All author discussed and interpreted the results. HR.L. and L.T. wrote the manuscript; XY.C., AP.L., and LH.T. edited the manuscript.

## Competing Interest Declarations

The authors declare no competing financial interests.

## Additional Information

See Supplementary Information.

# Figures

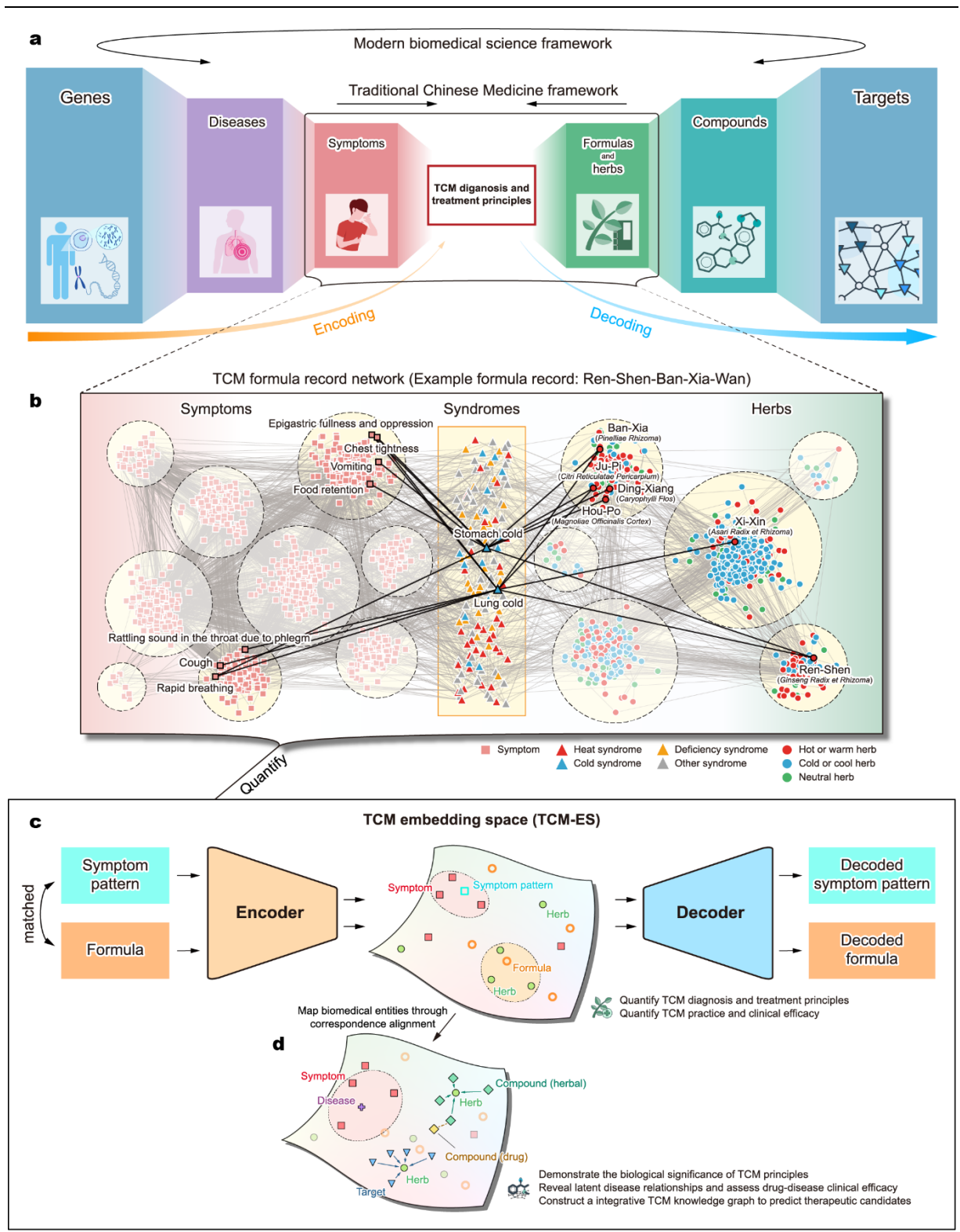

**Figure 1: Overview of the framework. (a)** Reductionism paradigm of modern biomedicine (upper path) and holistic paradigm of TCM (lower path). The biomedical reductionism path expands dimensionality at each layer, whereas TCM holistic principles directly map symptom patterns into herbal therapies. **(b)** A network representation of ancient and classical TCM formula records,

illustrating TCM diagnosis and treatment principles by linking patients' symptoms (squares), differentiated syndromes (triangles), and prescribed herbs (circles). Edge weights represent co-occurrence frequencies within the formula records, measured by normalized pointwise mutual information (see **Methods 6.1**). Nodes for herbs and syndromes are colour-coded according to TCM principles, distinguishing herb natures and syndrome pathogenesis (see **SI Text 1**). An example classical TCM formula record, *Ren-Shen-Ban-Xia-Wan*, demonstrates its associated symptoms, syndromes, and herbs. **(c)** An AI model with autoencoder architecture trained with ancient and classical formula records in (b), representing TCM entities (symptoms, symptom patterns, herbs and formulas) as measurable vectors to quantify the TCM diagnosis and treatment principles. The model's bottleneck representation is defined as the TCM embedding space (TCM-ES), which captures symptom pattern-herbal therapy mappings and enables quantification of TCM practice and clinical efficacy. **(d)** Biomedical entities were integrated into the TCM-ES through correspondence alignment: diseases were represented as sets of associated symptoms; herbal compounds and target proteins were embedded near their corresponding herbs; and drugs were embedded based on their chemical structural similarity to pre-embedded herbal compounds. The integrated TCM-ES allows for the exploration of TCM biological significance, identification of latent disease relationships, assessment of drug efficacy, and prediction of potential therapeutic candidates from a TCM perspective.

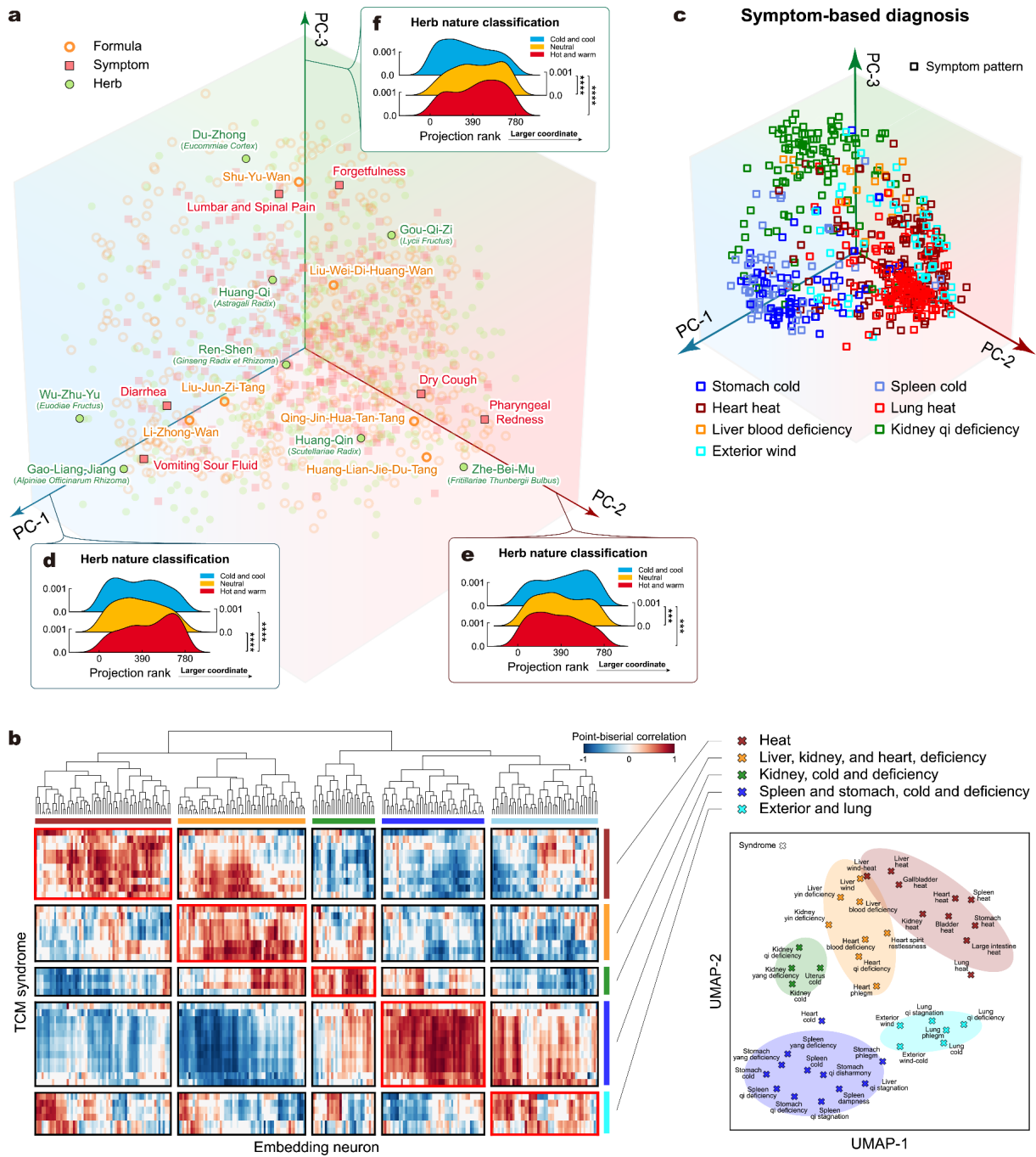

**Figure 2: The clinical experience derived TCM principles align with AI model encoding-decoding processes. (a)** The TCM Embedding Space (TCM-ES) learned by the model from TCM formula records, visualised via projection onto the first three principal components (PCs). Node colour and shape represent different types of TCM entities embedded in the space. **(b)** The alignment between TCM syndromes and the AI model embedding layer (bottleneck of the autoencoder) neurons. Note that syndrome information was excluded from training to ensure that the model learns objective mappings between symptom patterns and formulas. The tuning curves (left) show point-biserial correlations (PBC) between the activation of 256 neurons of the embedding layer (columns) and the presence of 40 most common syndromes (rows) in the formula data (**Methods 4.5**). Correlations are normalized to [0,1] for each row. The neurons were arranged by performing hierarchical clustering (left-top dendrogram) on column vectors. The TCM syndrome are visualised a in 2-D space (right) by performing UMAP on their associated symptom-and-herb vectors (**Methods 6.1**). Shaded regions indicate the confidence ellipses of syndrome clusters, with annotated key syndrome features. The diagonal block-wise pattern

suggests an alignment between syndromes and embedding neurons. **(c)** The distribution of seven representative TCM syndromes within the TCM-ES, with each scatter point representing a symptom pattern from a formula record, coloured by its differentiated syndrome. The continuous and well-clustered distribution of syndromes indicates that the TCM-ES provides a quantitative representation aligned with TCM syndrome differentiation. **(d-f)** The projection rank distributions for herbs categorised by TCM-defined natures (hot/warm, neutral, cold/cool) on the first 3 PCs. Higher ranks indicate larger coordinates on the PC (**Methods 5.1**). Note that information on herb natures was not used during model training. Significant differences in these distributions (Kolmogorov-Smirnov test. Significance levels: $p \leq 0.001(***), \leq 0.0001(****)$) demonstrates that herbs classified as having different natures exhibit distinct clustering patterns along different PCs of the TCM-ES.

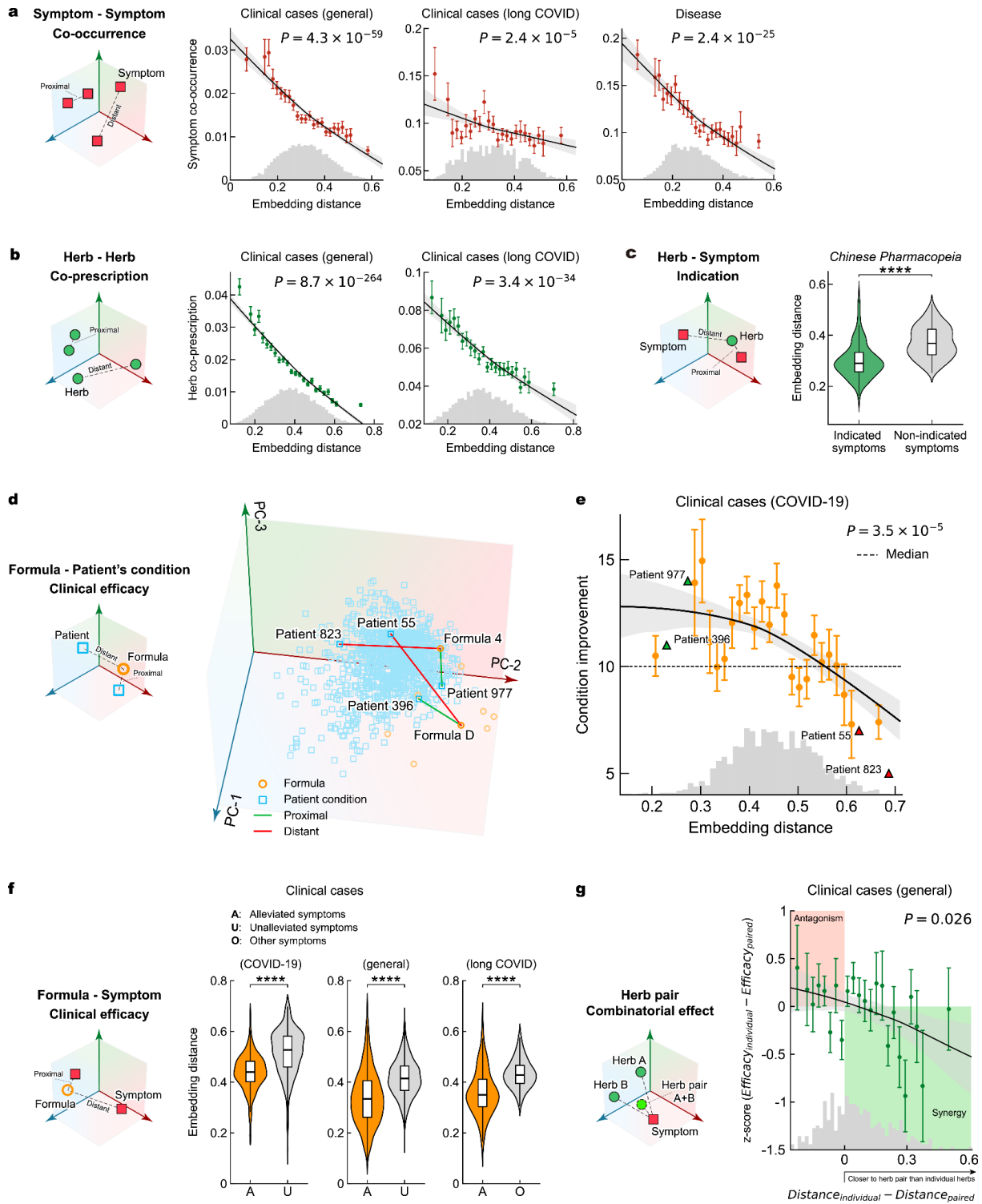

**Figure 3: The TCM-ES universally quantifies TCM practice and clinical efficacy.** (a) Pearson correlation coefficient (PCC) between embedding distances of symptom pairs and their co-occurrence frequencies (measured via the Jaccard Index) in clinical contexts (illustrated in the left panel), including general TCM cases (middle-left), long COVID cases (middle-right), and human diseases (right). Negative PCC indicates that symptoms closer in the TCM-ES co-occurred more frequently. $P$-values indicate the significance of correlations. Error bars represent standard errors for each bin. Black curves with shaded regions show the LOWESS fits with 5%-95% confidence intervals (see **Methods 7.5**). The bottom histograms show the distribution of embedding distances. (b) PCC between embedding distances of herb pairs and their co-prescription frequencies (illustrated in the left panel) in the general

TCM cases (middle) and long COVID cases (right). Herbs embedded closer in TCM-ES show higher co-prescription frequencies. **(c)** Comparison of embedding distances between herb-symptom pairs indicated in the *Chinese Pharmacopoeia* and random pairs. Officially indicated pairs exhibit significantly shorter distances ($p < 2.78 \times 10^{-64}$, Student *t*-test). **(d)** Visualization of COVID-19 patients' initial conditions (blue squares) and prescribed formulas (red circles) in the TCM-ES. The examples of condition-formula pairs demonstrate closer proximity for better matches (green lines) and greater distances for less matches (red lines). **(e)** Negative PCC between initial condition-formula embedding distances and follow-up symptom improvements, showing that better-matched pairs in TCM-ES correlated with greater improvements. **(f)** Comparison of embedding distances from prescribed formulas to alleviated symptoms against unalleviated or other symptoms (illustrated in the left panel). Alleviated symptoms show significant shorter distance to prescribed formulas (paired Student *t*-test) across COVID-19 cases (middle-left, $p = 1.24 \times 10^{-53}$), general TCM cases (middle-right, $p = 1.39 \times 10^{-247}$), and long COVID cases (right, $p = 3.87 \times 10^{-14}$). **(g)** The synergistic or antagonistic effects of herb pairs compared to individual herbs (illustrated in the left panel; **Methods 6.4**). *Z*-scores of clinical efficacy differences (y-axis) show negative PCC with the differences in embedding distance (x-axis), indicating that herb pairs embedded more proximal to symptoms exhibited superior clinical efficacy compared to individual herbs. Coloured regions highlight synergistic effects (green) and antagonistic effects (red).

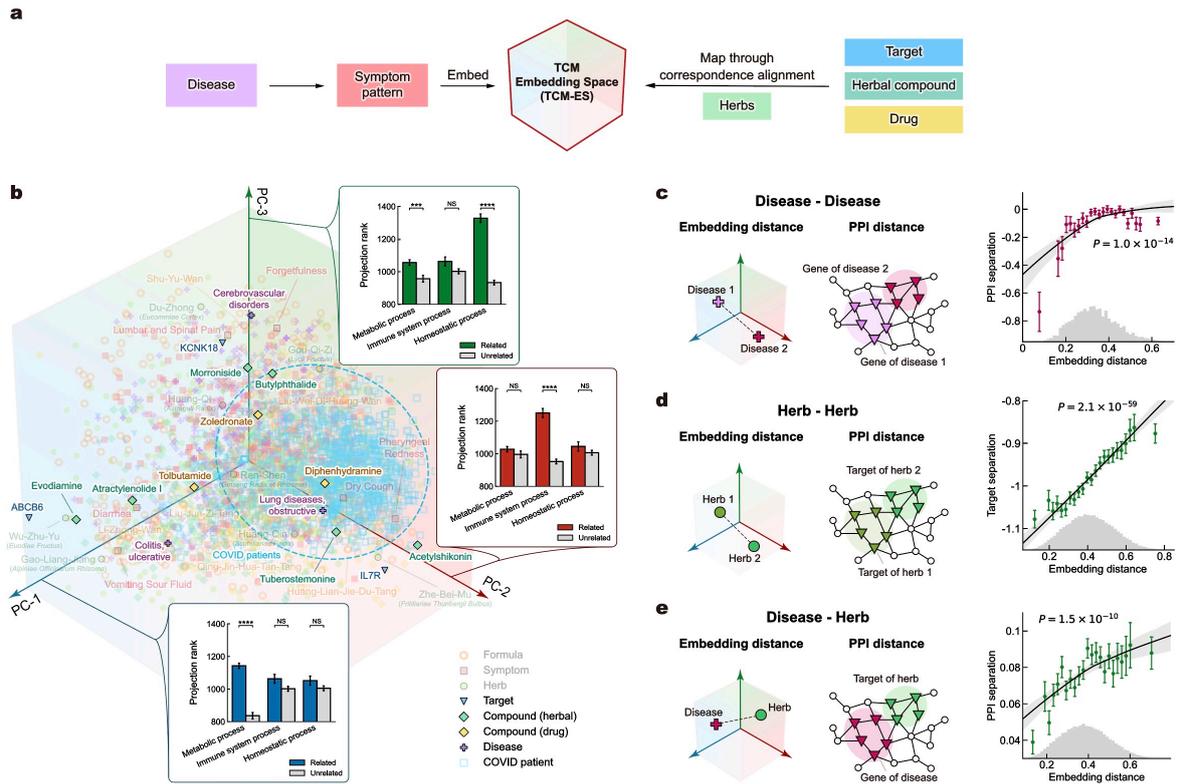

**Figure 4: The TCM-ES integrates biomedical entities and demonstrates biological significance.**
**(a)** Biomedical entities (diseases, herbal compounds, drugs, target proteins) were integrated into the TCM-ES based on the embeddings of their associated symptoms or herbs (see **Methods 4.6-4.7**). **(b)** The integrated TCM-ES and its biological significance. The TCM-ES is projected onto the first three principal components (PCs), The insets along the three PCs show projection rank differences for target proteins (see **Methods 5.1-5.2**), grouped by their relevance to key biological processes (e.g., metabolic, immune, and homeostatic). Significant rank differences (Student $t$-test. Significance levels: $p \leq 0.001(***), \leq 0.0001(****); \geq 0.05(NS)$) along specific PCs indicated their associations with key biological functions. **(c)** Positive Pearson correlation coefficient (PCC) between disease-disease embedding distances and their associated gene modules' separations (**Methods 6.5**) on the human protein-protein interaction (PPI) network. Disease pairs embedded closer in the TCM-ES exhibit smaller PPI separations (shorter distances) between their associated gene modules. **(d)** Positive PCC between herb-herb embedding distances and their target modules' separations on the PPI network. Herb pairs embedded closer in the TCM-ES tend to have target protein modules closer on the PPI network. **(e)** Positive PCC between disease-herb embedding distances and their associated gene/target modules' separations on the PPI network. Disease-herb pairs closer in the TCM-ES show smaller PPI separations between disease-associated genes and herb targets.

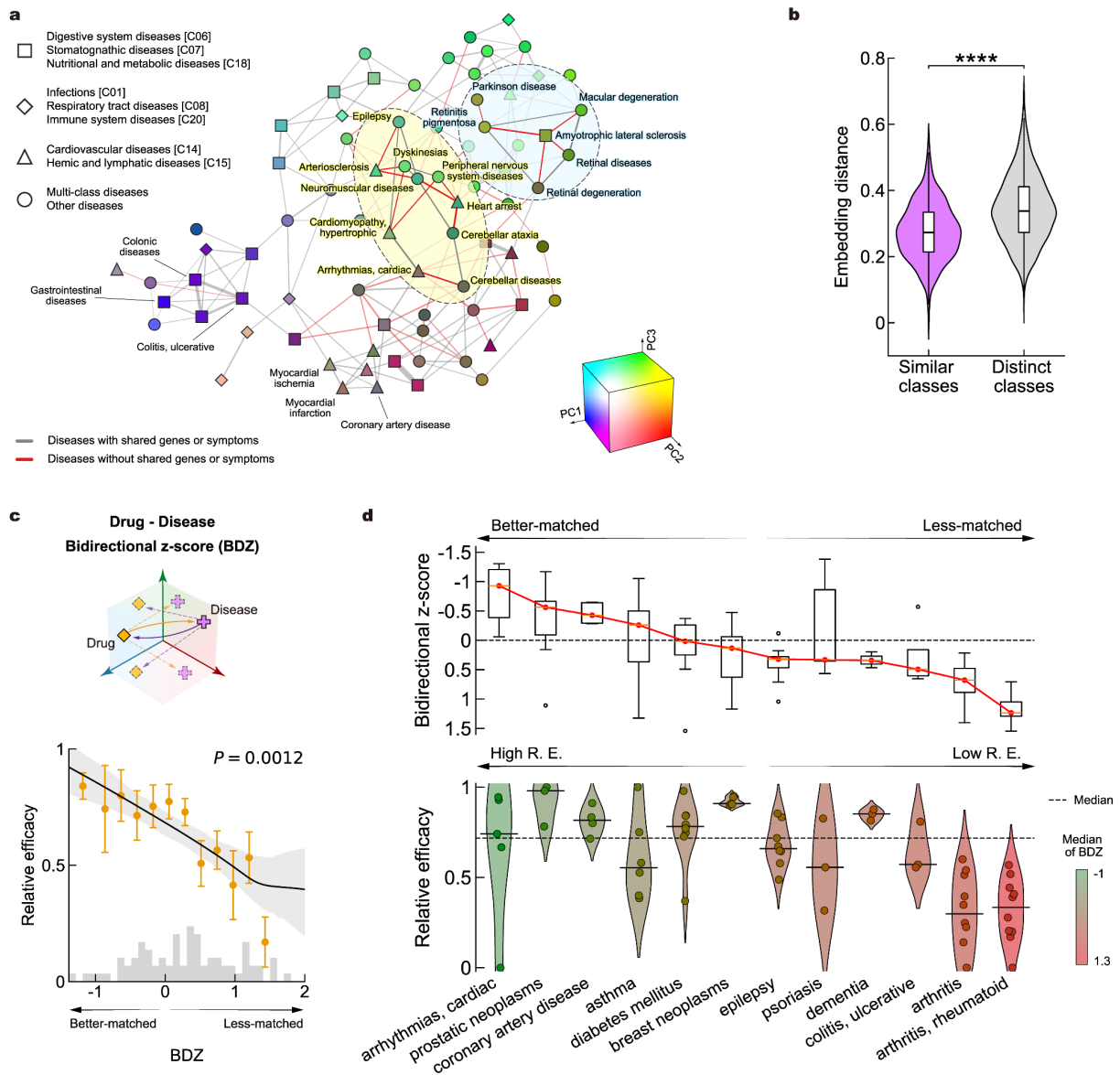

**Figure 5: The TCM-ES uncovers latent disease-disease relationships and assesses drug-disease clinical efficacy. (a)** A *K*-nearest neighbour (KNN) disease relationship network based on the TCM-ES. Each node represents a disease, with thicker edges indicating shorter embedding distances between diseases. Node colours reflect diseases' embedding projections onto the top three principal components of the TCM-ES, and node shapes correspond to disease MeSH categories. Grey edges connect disease pairs that share associated genes or symptoms, reproducing known disease relationships. Red edges highlight disease pairs without shared genes or symptoms, suggesting TCM-informed discovery of latent disease relationships. **(b)** Comparison of embedding distances from each disease to other diseases grouped by shared or distinct MeSH categories (see **Methods 3.1**). Diseases with shared MeSH categories exhibit significantly shorter embedding distances (paired Student *t*-test, $p = 4.15 \times 10^{-42}$). **(c)** Negative Pearson correlation coefficient between drug-disease Bi-directional *z*-scores (BDZ, lower value indicates closer embedding proximity, **Methods 6.6**) and Relative Efficacy (RE; **Methods 3.5**) scores derived from FDA Adverse Event Reporting System. Drugs that more closely match the diseases in the TCM-ES exhibit better clinical efficacy. **(d)** Top: BDZs for drugs associated with specific diseases in the TCM-ES. Bottom: RE scores for drugs treating these diseases. Solid short lines indicated median RE scores for each disease, while dashed lines denoted the overall median RE. Only diseases with at least three drugs reporting RE scores were included. Diseases such as *Cardiac Arrhythmias*

show better matching to indicated drugs and higher efficacy, while complex diseases, such as *Rheumatoid Arthritis*, exhibit weaker matching and lower efficacy.

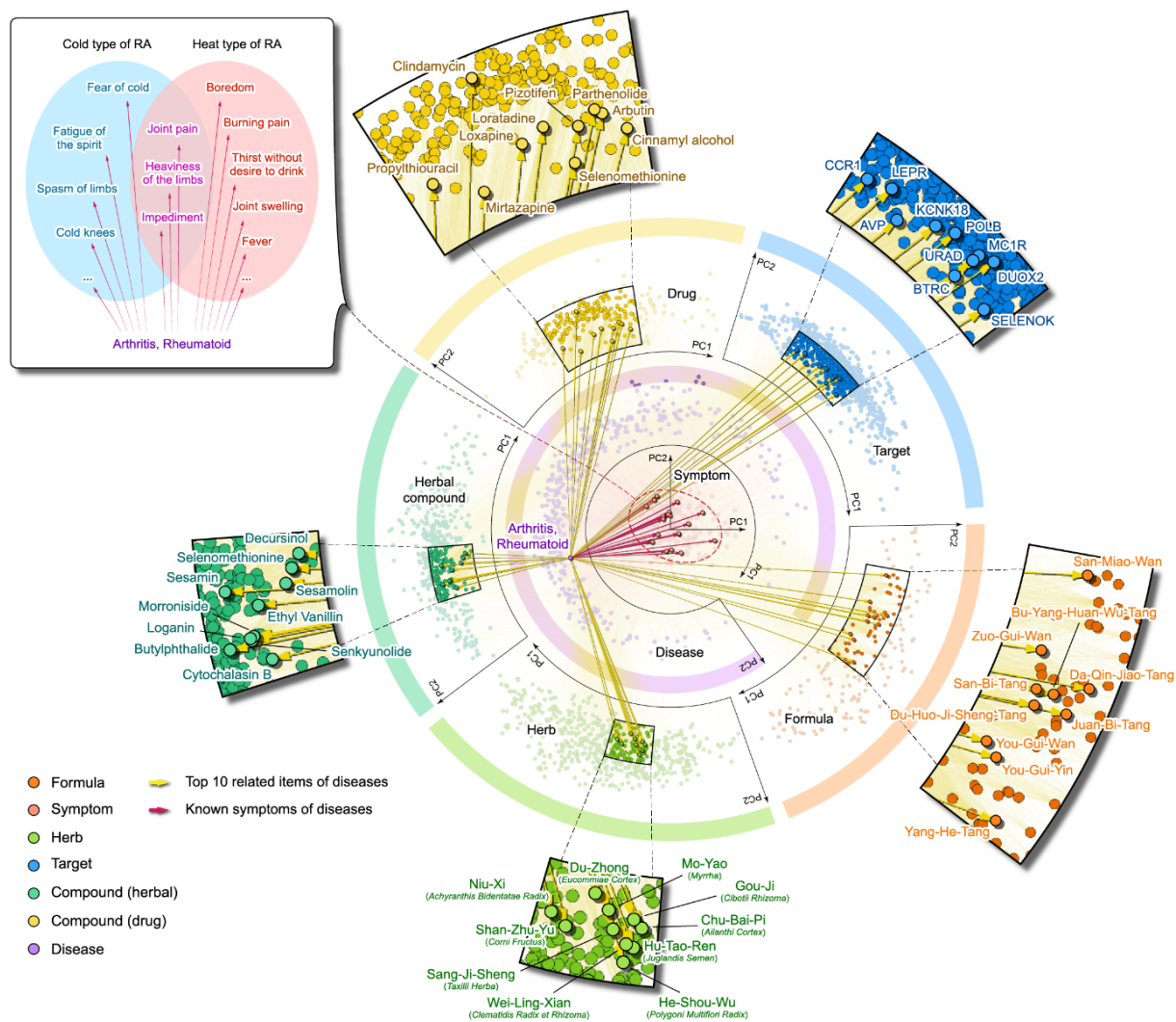

**Figure 6: A TCM-informed knowledge graph for predicting potential therapeutic candidates.** The knowledge graph was constructed based on the TCM-ES to identify therapeutic opportunities for human diseases. It comprises seven data types: symptoms, diseases, targets, drugs, herbal compounds, herbs, and formulas. Each data type is projected separately onto the first two principal components of the TCM-ES, with the coordinates representing their positions within the TCM-ES. Diseases were embedded based on associated symptoms (centre). Disease subtypes (e.g., Cold and Heat subtypes of *Rheumatoid Arthritis*) show distinct associated symptom clusters (left top, see **Extended Fig. E9** for more details of associated symptoms, genes, and indication drugs), reflecting the clinical heterogeneity. In the knowledge graph, each disease is connected to its 10 nearest embedding neighbours across data types, including proteins (molecular targets), herbs (empirical therapies), herbal compounds (therapeutic candidates), and drugs (repurposing opportunities). *Rheumatoid Arthritis* is presented as an example to illustrate the predictive connections described above, with some predictions supported by biological evidence or previous studies (**Extended Table E1**).